\documentclass[amsmath,superscriptaddress,amssymb,aps,twocolumn]{revtex4}

\usepackage{graphicx}
\usepackage{bm}

\usepackage[usenames]{color}

\begin{document}

\title{Experimental demonstration of topological error correction}

\author{Xing-Can Yao}
\affiliation{Hefei National Laboratory for Physical Sciences at Microscale and
Department of Modern Physics, University of Science and Technology
of China, Hefei, Anhui 230026, PR China}
\author{Tian-Xiong Wang}
\affiliation{Hefei National Laboratory for Physical Sciences at Microscale and
Department of Modern Physics, University of Science and Technology
of China, Hefei, Anhui 230026, PR China}
\author{Hao-Ze Chen}
\affiliation{Hefei National Laboratory for Physical Sciences at Microscale and
Department of Modern Physics, University of Science and Technology
of China, Hefei, Anhui 230026, PR China}
\author{Wei-Bo Gao}
\affiliation{Hefei National Laboratory for Physical Sciences at Microscale and
Department of Modern Physics, University of Science and Technology
of China, Hefei, Anhui 230026, PR China}
\author{Austin G. Fowler}
\affiliation{CQC2T, School of Physics, University of Melbourne, VIC 3010, Australia}
\author{Robert Raussendorf}
\affiliation{Department of Physics and Astronomy, University of British Columbia, Vancouver, BC, V6T 1Z1, Canada}
\author{Zeng-Bing Chen}
\affiliation{Hefei National Laboratory for Physical Sciences at Microscale and
Department of Modern Physics, University of Science and Technology
of China, Hefei, Anhui 230026, PR China}
\author{Nai-Le Liu}
\affiliation{Hefei National Laboratory for Physical Sciences at Microscale and
Department of Modern Physics, University of Science and Technology
of China, Hefei, Anhui 230026, PR China}
\author{Chao-Yang Lu}
\affiliation{Hefei National Laboratory for Physical Sciences at Microscale and
Department of Modern Physics, University of Science and Technology
of China, Hefei, Anhui 230026, PR China}
\author{You-Jin Deng}
\affiliation{Hefei National Laboratory for Physical Sciences at Microscale and
Department of Modern Physics, University of Science and Technology
of China, Hefei, Anhui 230026, PR China}
\author{Yu-Ao Chen}
\affiliation{Hefei National Laboratory for Physical Sciences at Microscale and
Department of Modern Physics, University of Science and Technology
of China, Hefei, Anhui 230026, PR China}
\author{Jian-Wei Pan}
\affiliation{Hefei National Laboratory for Physical Sciences at Microscale and
Department of Modern Physics, University of Science and Technology
of China, Hefei, Anhui 230026, PR China}

\date{\today}

\begin{abstract} 
Scalable quantum computing can only be achieved if qubits are manipulated fault-tolerantly.
Topological error correction---a novel method which combines topological quantum computing and quantum error 
correction---possesses the highest known tolerable error rate for a local architecture. 
This scheme makes use of cluster states with topological properties and requires only nearest-neighbour interactions. 
Here we report the first experimental demonstration of topological error correction with an eight-photon cluster state. 
It is shown that a correlation can be protected against a single error on any qubit,
and when all qubits are simultaneously subjected to errors with equal probability,
the effective error rate can be significantly reduced. 
This demonstrates the viability of topological error correction. 
Our work represents the first experimental effort to 
achieve fault-tolerant quantum information processing by exploring the topological properties of quantum states.
\end{abstract}

\maketitle

Quantum computers exploit the laws of quantum mechanics, and can solve many problems exponentially 
more efficiently than their classical counterparts \cite{Shor94,Grover97Search,Feynman82}. 
However, in the laboratory, the ubiquitous decoherence makes it notoriously hard to achieve 
the required high degree of quantum control. 
To overcome this problem, quantum error correction (QEC) has been invented \cite{Calderbank96,Steane96,Gottesman98}.
The capstone result in QEC, the so-called threshold theorem \cite{Knill05,Aliferis06}, 
states that as long as the error rate $p$ per gate in a quantum computer is smaller than a threshold value $p_c$, 
arbitrarily long and accurate quantum computation is efficiently possible. 
Unfortunately, most methods of fault-tolerant quantum computing with high 
threshold ($10^{-4}-10^{-2}$) require strong and long-range interactions \cite{Kitaev97,Knill05,Aliferis06}, 
and are thus difficult to implement. 
Local architectures are normally associated with much lower thresholds. 
For traditional concatenated codes on a 2D lattice of qubits with nearest-neighbour gates, 
the best threshold known to date \cite{Spedalieri09} is $2.02 \times10^{-5}$.

In such lattices, it is advantageous to employ topological 
error correction (TEC) \cite{Dennis02,Raussendorf06,wang11,Raussendorf07,Barrett10}
in the framework of topological cluster-state quantum computing.
This scheme makes use of the topological properties in three-dimensional (3D) cluster states,
which form an inherently error-robust ``fabric'' for computation. 
Local measurements drive the computation and, at the same time, implement the error correction.
Active error correction and topological methods are combined, 
yielding a high error threshold \cite{Raussendorf06,wang11} of 0.7\%--1.1\% and tolerating loss rates \cite{Barrett10} up to 24.9\%.
This leaves room for the unavoidable imperfections of physical devices, and 
makes TEC close to the experimental state of the art.
The 3D architecture can be further mapped onto a local setting in two spatial 
dimensions plus time \cite{Raussendorf07}, also with nearest-neighbour interactions only.
Two detailed architectures have already been proposed \cite{Stock08,Devitt08}. 
Note that a distinct and also important topological scheme has been proposed, in 
which quantum computation is driven by non-abelian anyons \cite{Nayak08,Wilczek90} 
and fault tolerance is achieved via passive stabilization afforded by a ground-state energy gap.

Some simple QEC codes have been experimentally demonstrated in nuclear magnetic 
resonance \cite{Cory98,Knill01}, ion traps \cite{Chiaverini04,Schindler11} and optical systems~ \cite{Chaoyang08,Aoki09}. 
However, the experimental realization of topological QEC methods still remains a challenging task. 
The state-of-the-art technology for generating multipartite cluster state is up to six photons, 
while great endeavor is still underway to create non-ablelian anyons for 
the topological quantum computing \cite{Nayak08,Wilczek90}. 
Here, we develop an ultra-bright entangled-photon source by utilizing an interferometric Bell-type synthesizer. 
Together with a noise-reduction interferometer, we generate a polarization-encoded eight-photon cluster state,
which is shown to possess the required topological properties for TEC.
In accordance with the TEC scheme, we measure each photon (qubit) locally. 
Error syndromes are constructed from the measurement outcomes, and one topological quantum correlation
is protected. We demonstrate: 
(1), if only one physical qubit suffers an error, the noisy qubit can be located and corrected, and 
(2), if all qubits are simultaneously subjected to errors with equal probability, 
the effective error rate is significantly reduced by error correction.
Therefore, we have successfully carried out a proof-of-principle experiment that
demonstrates the viability of {\em Topological Error Correction}---a central ingredient in 
topological cluster-state computing.

\begin{figure*}[t]
\begin{center}
\includegraphics[width=0.75\linewidth]{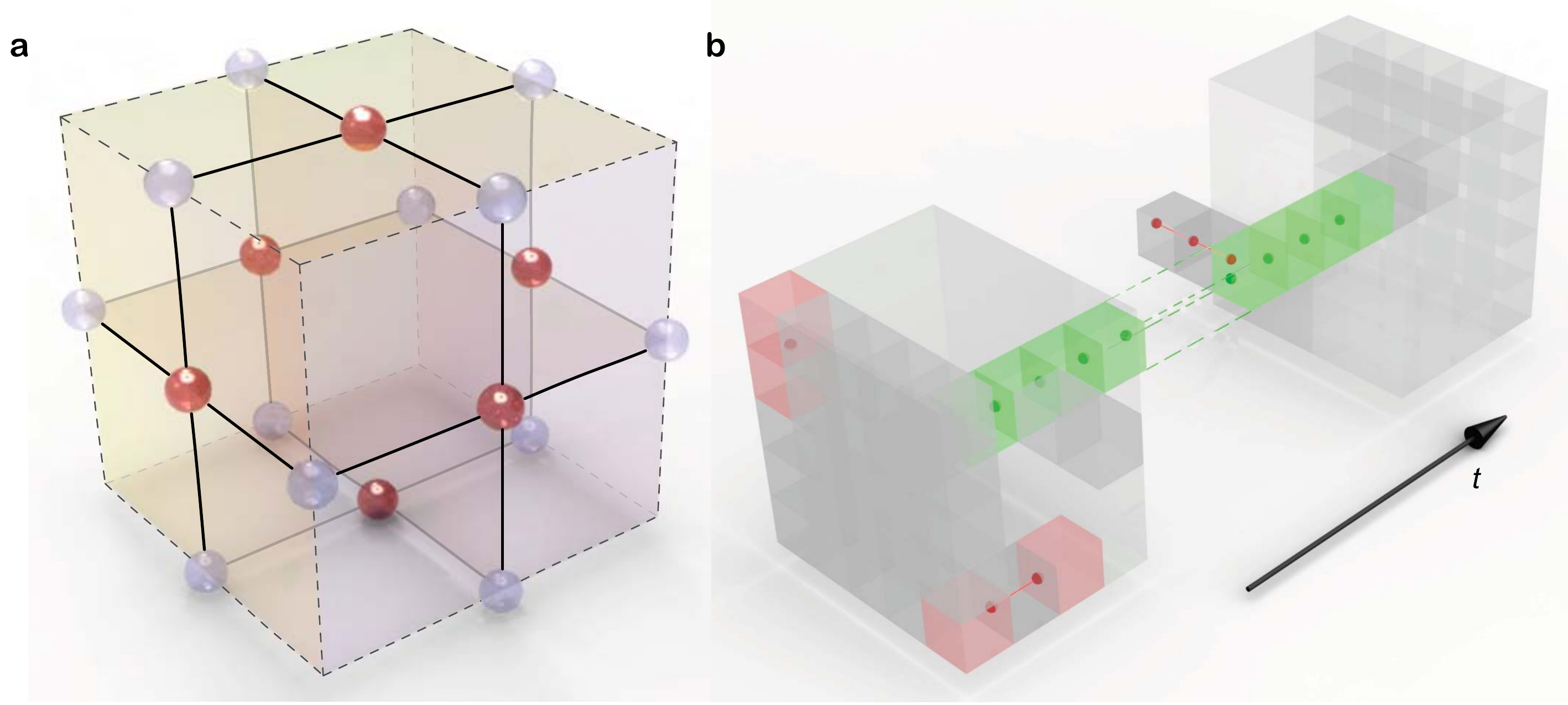}
\end{center}
\caption{\textbf{Topological cluster states.} The elementary lattice cell. 
Dashed lines represent the edges of the associated cell complex, while solid lines are for 
the edges of the interaction graph. 
Qubits live on the faces and the edges of the elementary cell. 
\textbf{b,} A larger topological cluster state of  $5\times5\times T$ cells. 
Green dots represent local $Z$ measurements, effectively removing these qubits from the cluster state 
and thereby creating a non-trivial topology capable of supporting a single correlation. 
Red dots represent $Z$ errors. Red cells indicate  $C_F=-1$ at the ends of error chains. 
One axis of the cluster can be regarded as simulating the ``circuit time'' $t$. 
The evolution of logical states from $t_1$ to $t_2$ is achieved by performing local $X$ measurements 
on all physical qubits between $t_1$ and $t_2$. }\label{fig:top}
\end{figure*}

\section*{Cluster states and quantum computing}
In cluster-state quantum computing \cite{Raussendorf01}, projective one-qubit measurements
replace unitary evolution as the elementary process driving a quantum computation. 
The computation begins with a highly entangled multi-qubit state, the so-called 
cluster state $\left|G\right\rangle$ \cite{Schlingemann01},
which is specified by an interaction graph $G$ and can be created from 
a product state via the pairwise Ising interaction over the edges in $G$.
For each vertex $i \in G$, one defines a stabilizer as $K_i : \equiv X_i \mathop  \otimes 
\limits_{e_{ij}} Z_j$, where the product is over all the interaction edges $e_{ij}$
connecting vertex $i$ to its neighbouring vertex $j$.  As usual, symbols $X_i$ and $Z_j$ denote 
the bit- and phase-flip Pauli operators, respectively, acting on qubits $i$ and $j$. 
State $\left|G\right\rangle$ is the unique joint eigenstate of a complete set of stabilizers $K_i$,
$K_i\left|G\right\rangle=\left|G\right\rangle$, for all the vertices $i \in G$. 

Cluster states in $d\geq 3$ dimensions are resources for universal fault-tolerant quantum computing \cite{Raussendorf06}. 
Therein, the TEC capability---shared with Kitaev's toric code \cite{Kitaev03,Dennis02} and 
the color code \cite{Bombin06}---is combined with the capability to process quantum information.

\section*{Topological error correction}
Quantum error correction and fault-tolerant quantum computing are possible with cluster states 
whenever the underlying interaction graph can be embedded in a 3D cell structure known as a cell 
complex \cite{Hatcher02}, which consists of volumes, faces, edges and vertices. 
Qubits live on the edges and faces of a cell complex. The associated interaction graph
connects the qubit on each face to the qubits on its surrounding edges via the interaction edges.
Consider the elementary cell complex in Fig.~1a, shown by the dashed lines, 
it has 1 cubic volume, 6 square faces, 12 edges, and 8 vertices. 
The interaction edges, specified by the solid lines, form an 18-qubit cluster state $\left|G_{18}\right\rangle$.
There are 6 face stabilizers $K_f \; (f=1,2,\cdots,6)$.
It follows that multiplication of these stabilizers cancels out all $Z$ operators in $K_f$ 
and thus yields a unit expectation value $\left\langle X_1X_2 \cdots X_6\right\rangle=1$. 
This leads to a straightforward but important observation that
despite the X-measurement on each individual face-qubit having random outcome $\pm1$, the product of 
all the outcomes on any {\it closed} surface $F$ is $+1$. Namely,  any closed surface has
the topological quantum correlation $C_F:\equiv\left\langle\otimes_{f\in F}X_f\right\rangle=1$.

A larger cell complex is displayed in Fig.~1b, which encodes and propagates a logical qubit. It consists of $5\times5\times T$ cells,
with $T$ specifying a span of simulated time $t$. A ``defect'' 
along the $t$ direction (shown as the line of green dots in Fig.~1b)
is first carved out via performing local $Z$ measurements. 
Then, the topological quantum correlation $C_{F_D}=1$ on a defect-enclosing closed surface,
combined with the boundary, is used to encode a logical qubit.  
The evolution of the logical state from $t_1$ to $t_2$ is achieved by local $X$ measurements 
on all other physical qubits between $t_1$ and $t_2$ (see Ref. \cite{Fowler08} for the details). 
Quantum computing requires a much larger cell complex and more defects, where quantum algorithms are realized by  
appropriate braiding-like manipulation of defects (a sketch for the logical CNOT gate is 
shown in Appendix). 

The quantum computation is possible precisely due to the topological quantum correlation $C_{F_D}=1$ 
on defect-enclosing {\em closed} surfaces $F_D$. 
The TEC capability arises from the $\textbf{Z}_2$ homology, a topological feature, of a sufficiently large 3D cell 
complex (see Appendix). For a given defect-enclosing closed surface $F_D$, 
there exist many homologically equivalent closed surfaces that represent the same topological correlation $C_{F_D}=1$.
This redundancy leads to the topological protection of the correlation \cite{Raussendorf06}.

Remarkably, in TEC it is sufficient to deal with $Z$ errors, 
because an $X$ error has either no effect if immediately before X measurements or is equivalent 
to multiple $Z$ errors.  Finally, as a measurement-based quantum computation, corrections
suggested by TEC are not applied to the remaining cluster state but rather 
to the classical outcomes of $X$ measurements.

\begin{figure}[t]
\begin{center}
\includegraphics[width=\linewidth]{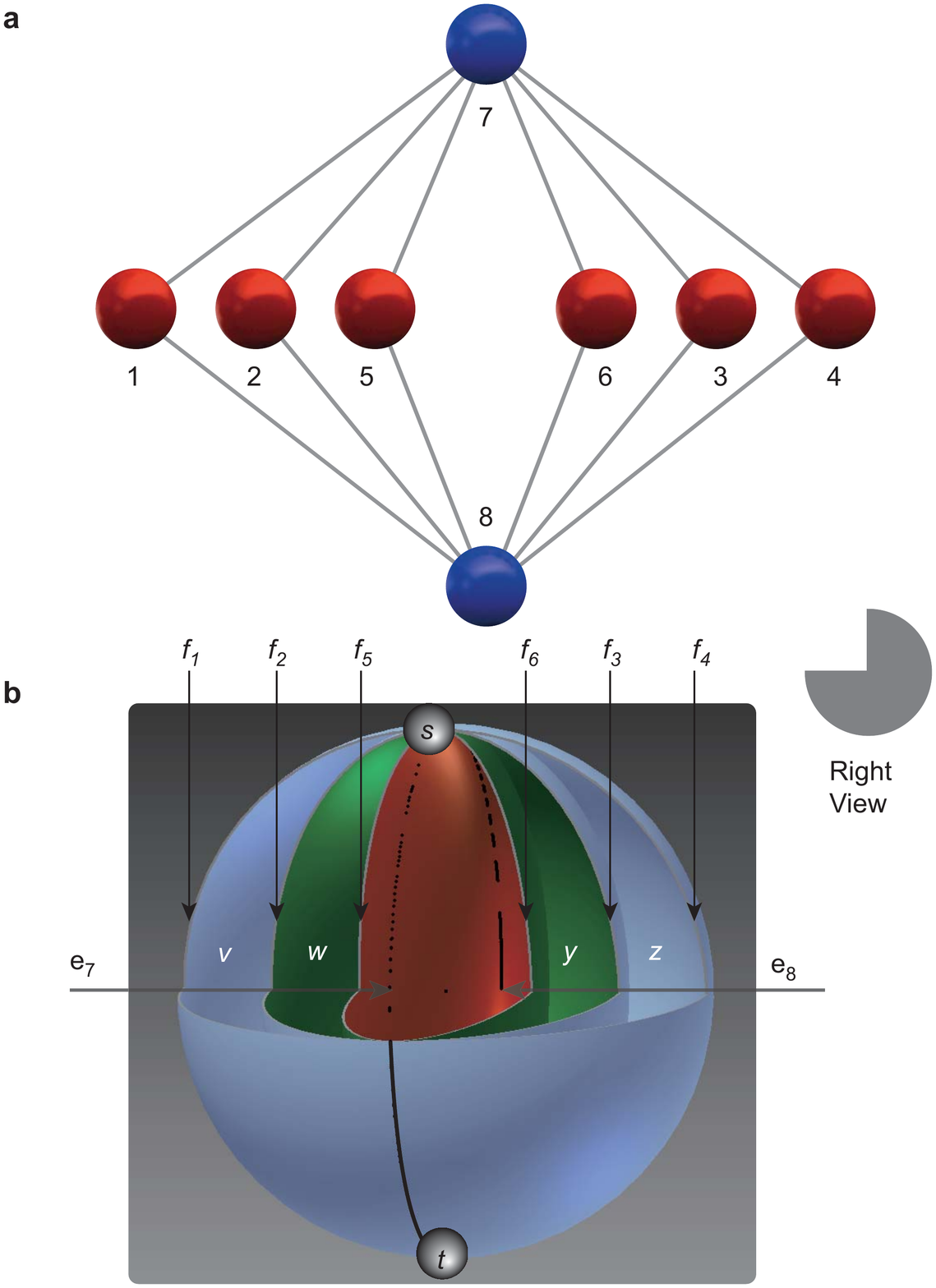}
\end{center}
\caption{\textbf{Cluster state  $|G_8\rangle$ and its cell complex. a} Interaction graph  $G_8$ of $|G_8\rangle$. 
\textbf{b,} The corresponding three-dimensional cell complex, with volumes $v,w,y,z$, faces $f_1,f_2,f_3,f_4,f_5,f_6$, 
edges  $e_7,e_{8}$, and vertices $s,t$ . The exterior and the center volume are not in the complex. 
For better illustration, the cell complex is cut open and the front up quarter is removed, c.f. ``right view''.}\label{fig:g8}
\end{figure}

\section*{Simpler topological cluster state}

The cell complex in Fig.~1b encodes a propagating logical qubit via 
one topological correlation  $C_{F_D}=1$, and is robust against a local $Z$ error. 
Unfortunately, it contains 180 physical qubits per layer, significantly beyond the reach of available techniques. 
We design a simpler graph state $\left|G_8\right\rangle$, shown in Fig.~2a, to mimic the cell complex of Fig.~1b.

The topological feature of $\left|G_8\right\rangle$ can be seen via its association with the 3D cell complex in Fig.~2b,
which consists of 4 elementary volumes $\left\{v,w,y,z\right\}$, 6 faces $\left\{f_1,f_2,f_3,f_4,f_5,f_6\right\}$, 
2 edges  $\{e_7,e_8\}$, and 2 vertices $\left\{s,t\right\}$.
All 6 faces have the same boundary $e_7 \cup e_8$, and any two of them forms a closed surface $F$.
The center volume is carved out, resembling the defect in Fig.~1b, and 
the to-be-protected topological correlation $C_{F_D}$ reads
\begin{equation}
C_{F_D}:\equiv\left\langle X_5 X_{6}\right\rangle=1.
\label{eq:cf}
\end{equation}
In this simple cell complex, the topological correlation $C_{F_D}=1$ is already multiply encoded,
represented by any expectation $\langle X_i X_j \rangle$ with $i \in \{1,2,5\}$ and $j \in \{3,4,6\}$.
Moreover, there exist four other closed surfaces without enclosing the defect,
corresponding to the boundary of volumes $v$, $w$, $y$, $z$, respectively.
The ``redundant'' topological correlations are
\begin{equation}
 \left\langle X_1 X_2\right\rangle=\left\langle X_2 X_5\right\rangle=\left\langle X_3 X_6\right\rangle
 =\left\langle X_3 X_4\right\rangle = 1 \; ,
\label{eq:topcorrelation}
\end{equation}
and can be used as error syndromes in TEC. As shown in Table 1, 
a single $Z$ error on any physical qubit can be located and corrected.

\begin{table}
\caption{Location of a Z error in  $\vert G_8\rangle$ and the syndromes  $C_{12}=\langle X_1X_2\rangle$ etc.}
\centering\begin{tabular}{c c c c c}
\hline\hline
$Z_{\mbox{error}}$& $C_{12}$& $C_{25}$& $C_{36}$& $C_{34}$ \\
\hline
1 &-1& 1& 1& 1 \\
2 &-1& -1& 1& 1 \\
3 & 1 & 1 & -1 & -1 \\
4 & 1 & 1 & 1 & -1 \\
5 & 1 & -1 & 1 & 1 \\
6 & 1 & 1 & -1 & 1 \\
\hline
\end{tabular}
\label{table:error_syndromes}
\end{table}

Therefore, from the aspect of TEC capability, the  cluster state $\left|G_8\right\rangle$ is analogous to 
the cell complex in Fig.~1b. They protect one topological correlation and are robust against a single $Z$ error,
albeit the cell complex in Fig.~2b is too small to propagate a logical qubit 
(see Appendix for detailed discussion). 

\begin{figure*}[t]
\begin{center}
\includegraphics[width=\linewidth]{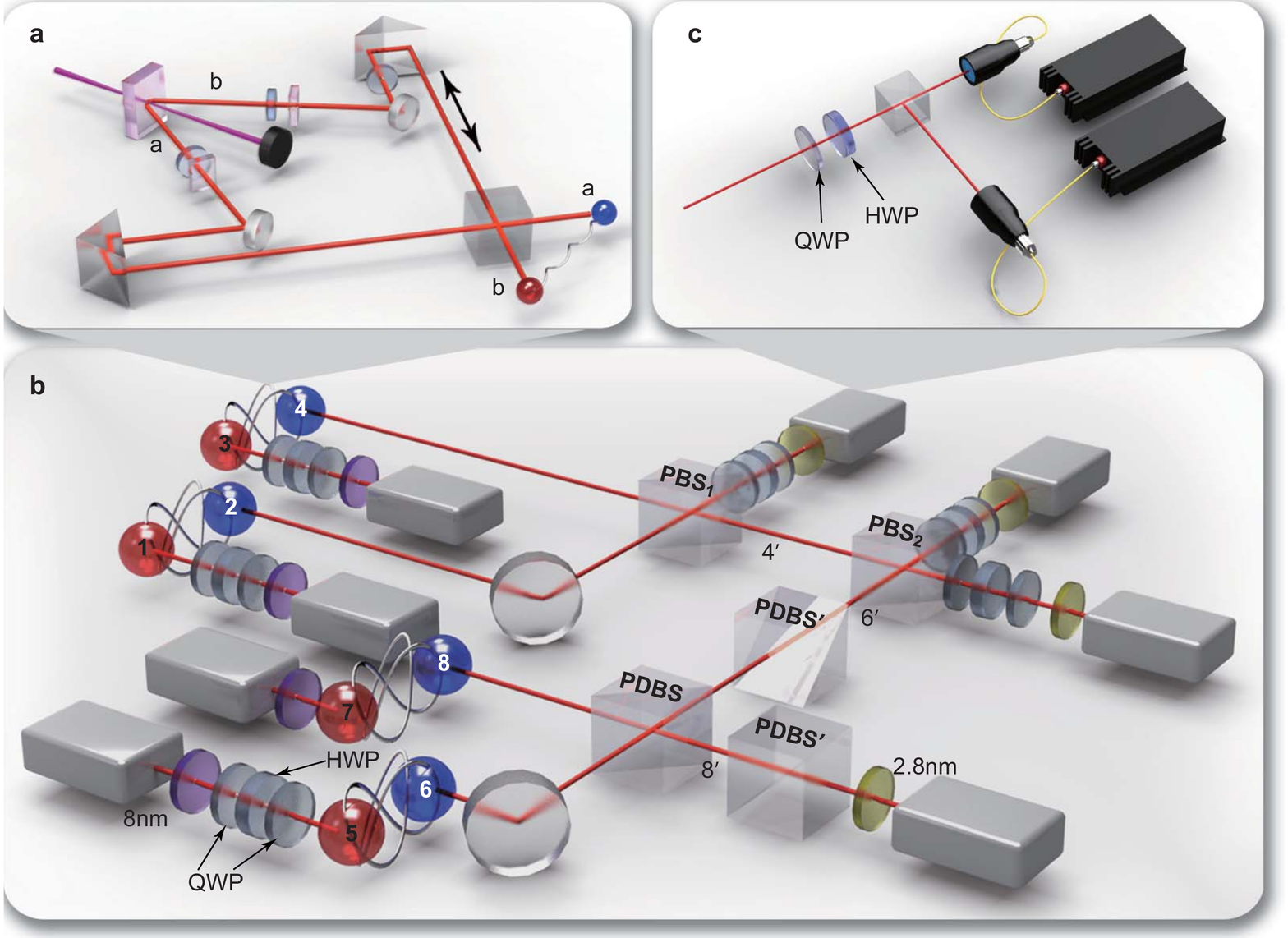}
\end{center}
\caption{\textbf{Experimental setup for the generation of the eight-photon cluster state and the demonstration of topological error correction. a} Creation of ultra-bright entangled photon pairs. 
An ultraviolet laser pulse passes through a 2~mm nonlinear BBO crystal, creating an entangled photon pair 
by parametric down conversion with $\rho  = \frac{1}{2}(\left| {H_a^o } \right\rangle \left| {V_b^e }
\right\rangle \left\langle {V_b^e } \right|\left\langle {H_a^o }
\right| + \left| {V_a^e } \right\rangle \left| {H_b^o }
\right\rangle \left\langle {H_b^o } \right|\left\langle {V_a^e }
\right|)$, where $o$ and $e$ indicate the polarization with respect to the $V$-polarized pump. After both photons pass through compensators including a 45$^0$ HWP and a 1~mm BBO crystal, 
one of the photons' polarizations is rotated by another 45$^0$ HWP. 
Then we re-overlap  the two  photons on a PBS, creating an entangled photon pair 
with $|\phi_{ab}\rangle=\frac{1}{{\sqrt 2 }}(\left| {H
} \right\rangle \left| {H } \right\rangle  + e^{i\varphi }
\left| {V } \right\rangle \left| {V } \right\rangle )\otimes\left|e_a\right\rangle\left|o_b\right\rangle$. 
\textbf{b,} In order to create the desired cluster state, we combine photons from path 6 and 8 at PDBS and 
let each photon pass through another PDBS', resulting a controlled-phase operation between photon 6 and 8. 
Meanwhile photon 2 and photon 4 are interfered on PBS$_1$. In the end, photon 4' and photon 6' are overlapped on PBS$_2$. 
Upon a coincidence detection, we create the eight-photon cluster state (\ref{eq:g8}) for topological error correction.
\textbf{c,} Polarization analyzer for each individual photon, containing a QWP, an HWP, a PBS and two single-mode 
fibre-coupled single-photon detectors.}\label{fig:setup}
\end{figure*}

\begin{figure*}[t]
\begin{center}
\includegraphics[width=\linewidth]{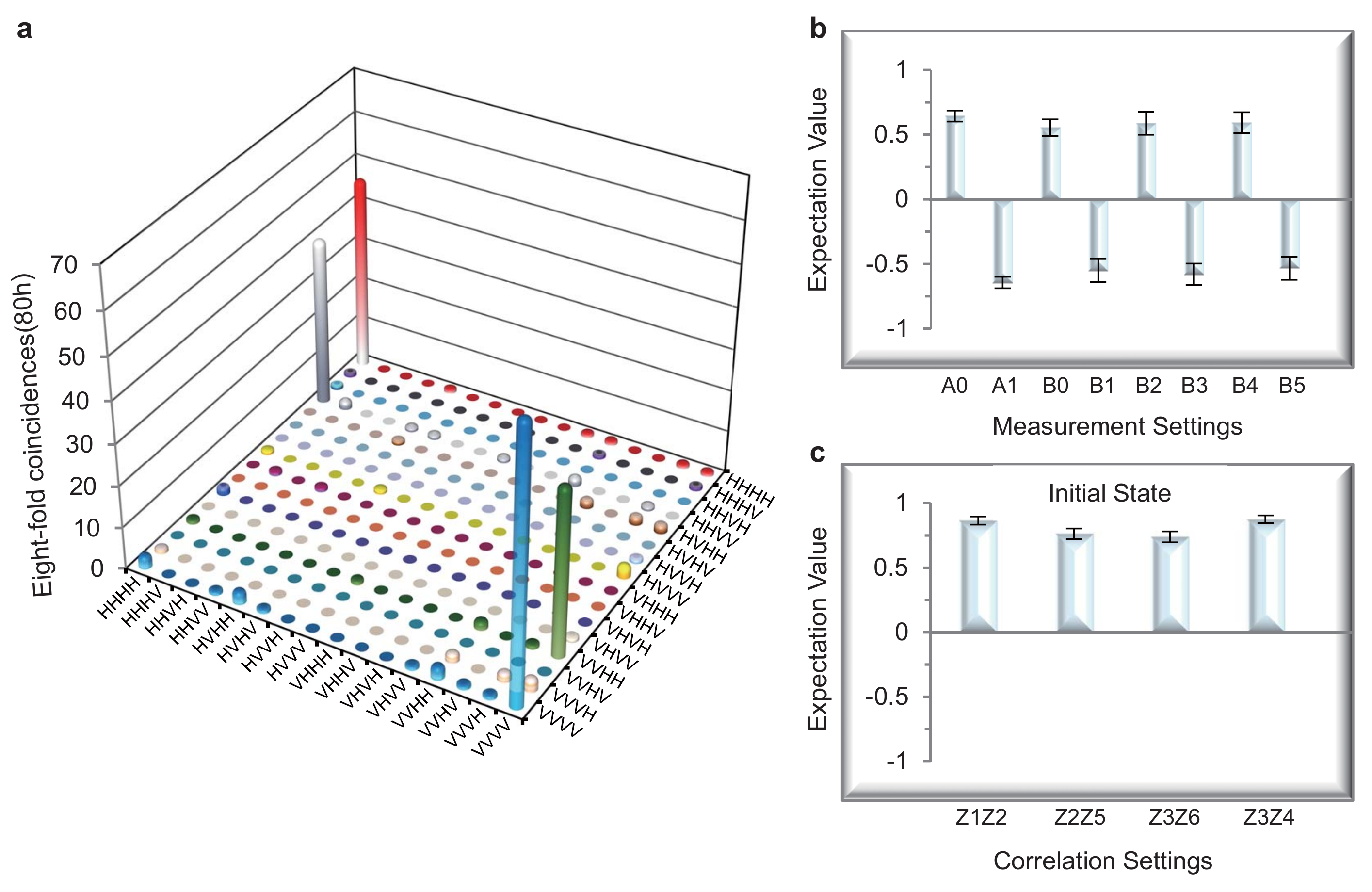}
\end{center}
\caption{\textbf{Experimental results for the created eight-photon cluster state. a,} Measured 
eight-fold coincidence in $|H\rangle/|V\rangle$ basis. \textbf{b,} The expectation values for different witness 
measurement settings. From left to right, the measurement settings 
are $\mbox{A}_0=(|H\rangle\langle H|^{\otimes 6}-|V\rangle\langle V|^{\otimes 6})_{1-6}\mbox{X}_7\mbox{X}_8$, 
$\mbox{A}_1=(|H\rangle\langle H|^{\otimes 6}-|V\rangle\langle V|^{\otimes 6})_{1-6}\mbox{Y}_7\mbox{Y}_8$, 
and $\mbox{B}_i =\mbox{M}_i^{\otimes6}(|H\rangle \langle H|^{\otimes 2}-|V\rangle \langle V|^{\otimes 2})_{78}$ with $i=0,\cdots,5$. 
The measurement of each setting takes 50 hours for the first two settings and 30 hours for the remainings.
\textbf{c,} Correlations for initial state without any engineered error. The error bars represent one standard deviation, 
deduced from propagated poissonian counting statistics of the raw detection events.
}\label{fig:hvbasis}
\end{figure*}

\section*{Preparation of the eight-photon cluster state}

In our experiment, the desired eight-qubit cluster state is created using spontaneous parametric down-conversion 
and linear optics. 
The first step is to develop an ultra-bright and high-fidelity entangled-photon source. As shown in Fig.~3a,
an ultraviolet mode-locked laser pulse (915 mW) passes through a $\beta$-barium borate (BBO) crystal, 
generating a pair of polarization-entangled photons in the state $|\phi\rangle=\left(|HH\rangle+
|VV\rangle\right)/\sqrt{2}$. By an interferometric Bell-state synthesizer \cite{Yao11}, 
photons of different bandwidths (shown by red and blue dots in Fig.~3a, respectively) are guided through separate paths. 
This disentangles the temporal from the polarization information. 
In contrast to the conventional narrow-band filtering technique, there is no photon-loss problem,
and thus an ultra-high brightness is achieved.
Four pairs of such entangled photons are prepared and labelled as  1-2, 3-4, 5-6 and 7-8  in Fig.~3b. 
Then, we generate two graph states, each of four photons. 
The first one is a four-photon GHZ state $\left(|H^{\otimes 4}\rangle_{1\mbox{-}4}\rangle+
|V^{\otimes 4}\rangle_{1\mbox{-}4}\right) /\sqrt{2}$, obtained by superposing photon 2 and photon 4
on a polarizing beam-splitter (PBS$_1$) which transmits $H$ and reflects $V$ polarization. 
Meanwhile, photon 6 and photon 8 are interfered on a polarization-dependent beam-splitter (PDBS)
and then separately pass through two PDBSs. 
The former has transmitting probabilities $T_H=1,\,T_V=1/3$ and the latter have $T_H=1/3,\, T_V=1$.
The combination of these three PDBSs acts as a controlled-phase gate \cite{Hofmann02,Kiesel05}.
With a success probability of 1/9, one has the twofold coincidence in path 6' and 8',
yielding a four-photon cluster state \cite{Kiesel05} $[|HH\rangle_{56}\left(|HH\rangle_{78}+|VV\rangle_{78}\right)+
|VV\rangle_{56}\left(|HH\rangle_{78}-|VV\rangle_{78}\right)] /2$.
Finally, photon 4' and photon 6' are superposed on PBS$_2$.
When eight photons come out of the output ports simultaneously, one obtains an entangled eight-photon cluster state:
\begin{widetext}
\begin{equation}
  |\psi\rangle=\frac{1}{2}\left[|H^{\otimes 6}\rangle_{1\mbox{-}6}\left(|HH\rangle_{78}+
  |VV\rangle_{78}\right)+|V^{\otimes 6}\rangle_{1\mbox{-}6}\left(|HH\rangle_{78}-|VV\rangle_{78}\right)\right].
\label{eq:g8}
\end{equation}
\end{widetext}
This is exactly the cluster state  $\left|G_8\right\rangle$ shown in Fig.~2a 
under Hadamard operations $H^{\otimes 8}$  on all qubits. 
Note that the photons, which are interfered on the PBSs or at the PDBS, have the same bandwidth,
and a star topology of the eight-photon interferometer leads to an effective noise-reduction.

\begin{figure*}[t]
\begin{center}
\includegraphics[width=\linewidth]{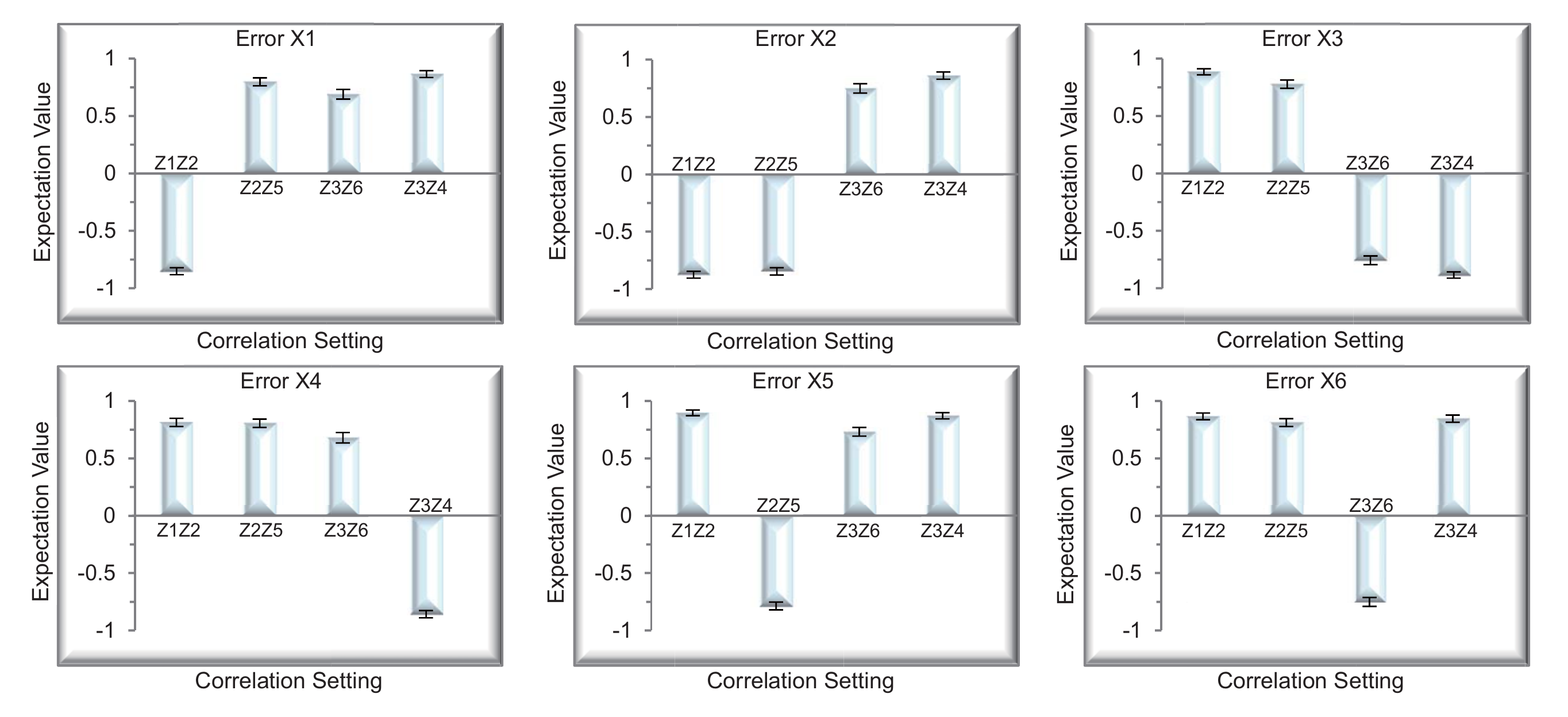}
\end{center}
\caption{\textbf{Experimental results of syndrome correlations for topological error correction.} 
Only one qubit is subjected to an $X$ error in each sub-figure. 
The measurement for  each error setting takes about 80 hours.
The error bars represent one standard deviation, 
deduced from propagated poissonian counting statistics of the raw detection events.}\label{fig:syndrome}
\end{figure*}

To ensure good spatial and temporal overlap, the photons are also spectrally filtered,
with $\Delta\lambda_{\tiny{\mbox{FWHW}}}= 8$ nm for 1-3-5-7 and $\Delta\lambda_{\tiny{\mbox{FWHW}}}= 2.8$ nm 
for 2-4-6-8, and coupled by single-mode fibres. 
We obtain an average two-fold coincidence count of about  $3.4\times10^5$ /s 
and a visibility of $\sim$94\% in the $|H\rangle/|V\rangle$ as well as in the $|+\rangle/|-\rangle$  basis, 
where $|\pm\rangle=\frac{1}{\sqrt 2}\left(|H\rangle\pm|V\rangle\right)$.  
Fine adjustments of the delays between the different paths are tuned to ensure that 
all the photons arrive at the PBSs and PDBS simultaneously. 

Measurement is taken for each individual photon by a polarization analyzer, which contains a combination of a QWP, a HWP 
and a PBS, together with two single-mode fibre-coupled single-photon detectors in each output of the PBS (see Fig.~3c). 
The complete set of the 256 possible combinations of eight-photon coincidence events is registered by 
a home-made FPGA-based programmable coincidence logic unit.
We obtain an eight-fold coincidence rate of 3.2 per hour. 
Based on the measurements for the 256 possible polarization combinations in the $|H\rangle/|V\rangle$ basis (Fig.~4a), 
we obtain a signal-to-noise ratio of about 200:1, defined as the ratio of the average of the desired components to 
that of the non-desired ones. This indicates the success of preparing the desired eight-photon cluster state.

To more precisely characterize the cluster state, we use the entanglement-witness method to determine its fidelity. 
For this purpose, we construct a witness which allows for the lower bound on the state fidelity 
and requires only eight measurement settings (see Appendix):
\begin{widetext}
\begin{eqnarray}
\mathcal {W}_8 &=& \frac{1}{2}-\left(|\psi\rangle\langle\psi|-|\psi^{\prime}\rangle\langle\psi^{\prime}|\right) \nonumber \\
&=& \frac{1}{2}-\left[\frac{1}{4}\left(|H\rangle\langle H|^{\otimes 6}-|V\rangle\langle V|^{\otimes 6}\right)_{1\mbox{-}6}\otimes\left(\mbox{X}_7\mbox X_8-\mbox Y_7\mbox Y_8\right) +\right. \nonumber \\
&& \left.\frac{1}{12}\left(\sum^{5}_{{k=0}}(-1)^k \mbox{M}_k^{\otimes 6}\right)_{1\mbox{-}6}\otimes\left(|H\rangle\langle H|^{\otimes 2}-|V\rangle\langle V|^{\otimes 2}\right)_{78}\right],
\label{eq:witness}
\end{eqnarray}
\end{widetext}
where $\langle \psi^{\prime}|\psi\rangle=0$ and $\mbox{M}_k=\left[\mbox{cos}(\frac{k\pi}{6})\mbox{X}+\mbox{sin}(\frac{k\pi}{6})\mbox{Y}\right]$. 
The results are shown in Fig.~4b, which yields the witness $\langle W\rangle=-0.105\pm0.023$, 
which is negative by 4.5 standard deviations. 
The state fidelity is $F>\frac{1}{2}- \langle W \rangle=0.605\pm0.023$.
The presence of genuine eight-photon entanglement is confirmed.

\section*{Experimental topological error correction}

 Given such a cluster state, topological error correction is implemented using a series of single-qubit 
 measurements and classical correction operations. 
 In the laboratory, operations are performed on state~(\ref{eq:g8}), 
 differing from $\left|G_8\right\rangle$ in Fig.~2a by Hadamard operation $H^{\otimes 8}$. 
 Therefore, the to-be-protected correlation  $\left\langle X_5 X_{6}\right\rangle$ in Eq. (\ref{eq:cf}) 
 corresponds to $\left\langle Z_5 Z_{6}\right\rangle$  in the experiment; 
 the same applies to the syndrome correlations (\ref{eq:topcorrelation}). 
 Meanwhile, $X$ errors are engineered instead of $Z$ errors.

 In the experiment, the noisy quantum channels on polarization qubits are engineered by one HWP sandwiched 
 with two QWPs, which are set at 90 degrees. By randomly setting the HWP axis to 
 be oriented at $\pm \theta$  with respect to the horizontal direction, the noisy quantum channel can be 
 engineered with a bit-flip error probability of  $p=\mbox{sin}^2(2\theta)$. 

 We first study the case that only a single $X$ error occurs on one of the six photons $\{ 1, \cdots, 6 \}$. 
 The syndrome correlations are measured, and the results are shown in Fig.~5. 
 For comparison, we also plot the correlations without any engineered error in Fig.~4c. 
 Indeed, one can precisely locate the physical qubit undergoing an $X$ error.

\begin{figure}[t]
\begin{center}
\includegraphics[width=\linewidth]{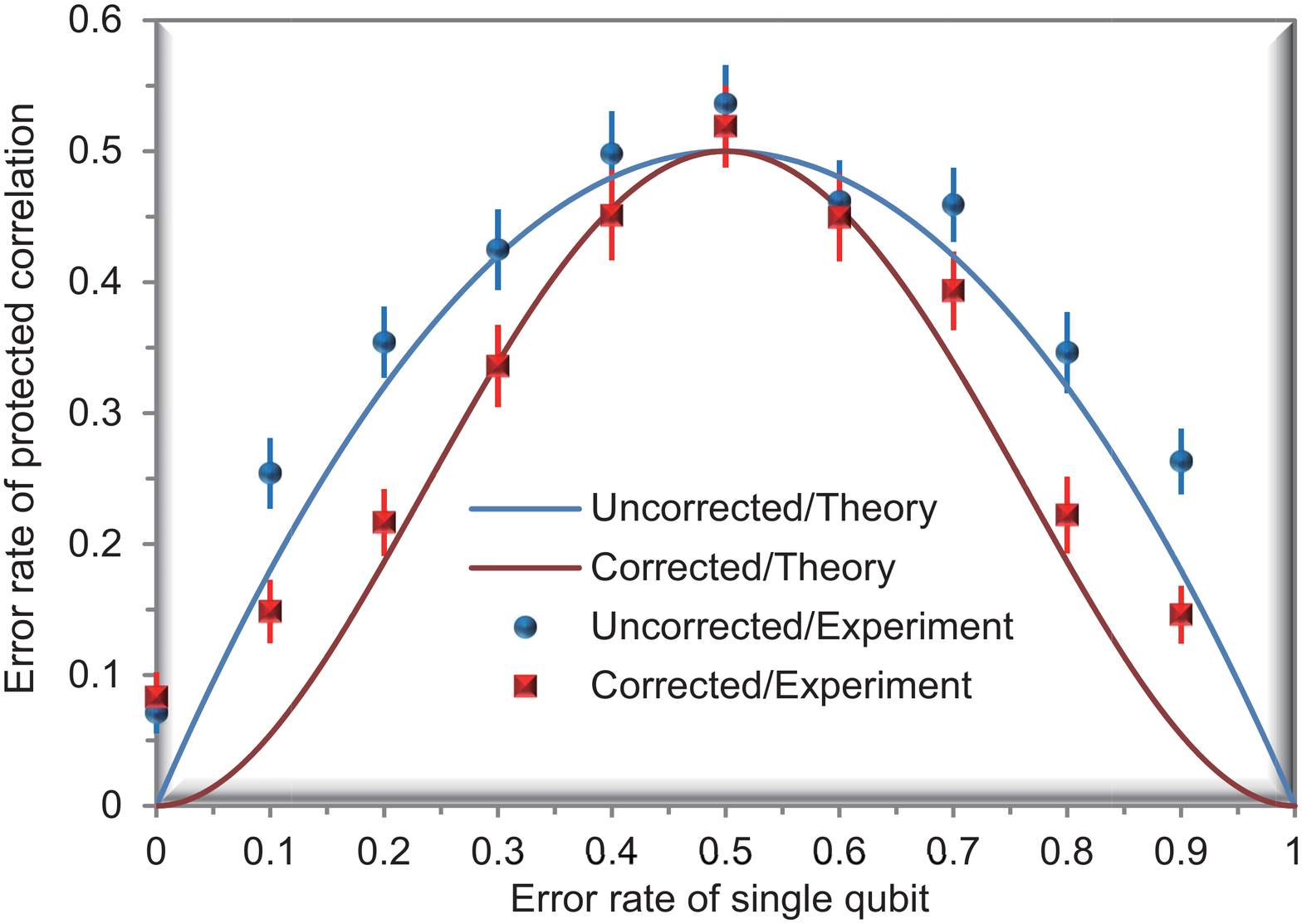}
\end{center}
\caption{\textbf{Experimental results of topological error correction.} All physical qubits are simultaneously subject to an $X$ error with equal probability ranging from 0 to 1. The blue round dots (blue lines) represent the experimental (theoretical) values of the error rate for the protected correlation without TEC, and the red square dots (red lines) are for the error rate with TEC. The agreement between the experimental and the theoretical results clearly demonstrates the viability of TEC. The measurement of each data point takes 80 hours. The error bars represent one standard deviation, deduced from propagated poissonian counting statistics of the raw detection events.}\label{fig:ted}
\end{figure}

 We then consider the case that all the six photons are simultaneously subject to a random $X$ error with equal 
 probability $0<p<1$, and study the rate of errors, $\langle Z_5 Z_6 \rangle =-1$, for the topological 
 quantum correlation $\langle Z_5Z_6 \rangle$. 
 Without error correction, the error rate of correlation $\langle Z_5Z_6 \rangle$ is $P=1-(1-p)^2-p^2$.
 With error correction, the residual error becomes
 \begin{widetext}
 \begin{equation}
  P=1-\left[(1-p)^6+p^6\right]-\left[6p(1-p)^5+6(1-p)p^5\right]-\left[9p^2(1-p)^4+9(1-p)^2p^4\right] \; .
  \label{eq:p}
 \end{equation}
 \end{widetext}
 For small $p$, the residual error rate after error correction is significantly reduced 
 as compared to the unprotected case. 
 As shown in Fig.~6, the experimental results are in good agreement with these theoretical predictions.
 Considerable improvement of the robustness of the $\langle Z_5Z_6 \rangle$ correlation can be seen both in theory and in practice.

 In the experiment, the whole measurement takes about 80 days.
 This requires an ultra stability of our setup.  
 The imperfections in the experiment are mainly due to the undesired components in the $|H\rangle/|V\rangle$ basis,
 arising from higher-order emissions of entangled photons, and the imperfect photon overlapping at the 
 PBSs and the PDBS.
 In spite of these imperfections, the viability of TEC is clearly demonstrated in the experiment.

\section*{Discussion}

 In the current work, we have experimentally demonstrated TEC with an eight-photon cluster state. 
 This state represents the current state-of-the-art for preparation of cluster states in any qubit system 
 and is of particular interest in studying multipartite entanglement and quantum information processing. 
 The scalable construction of cluster states in the future will require further development of high-efficiency 
 entanglement sources and single-photon detectors \cite{Obrien07}. Recent results have shown that if the product of 
 the  number-resolving detector efficiency and the source efficiency is greater than 2/3, efficient linear 
 optical quantum computation is possible \cite{Varnava08}. Solid technical progress towards this goal has been made 
 such as deterministic storable single-photon sources \cite{Shuai06} and photon-number-resolving detectors \cite{Kardynal08}.
 This work represents the first experimental demonstration of TEC, an important step towards fault-tolerant 
 quantum computation. In the scheme, given sufficient qubits and physical error rates below 0.7\%--1.1\%, arbitrary 
 quantum computations could be performed arbitrarily reliably. The high threshold error rate is especially remarkable 
 given that only nearest neighbour-interactions are required. Due to these advantages, TEC is especially well-suited for 
 physical systems geometrically constrained to nearest-neighbour interactions, such as quantum dots \cite{Press08}, 
 Josephson junction qubits \cite{Hime06}, ion traps \cite{Hensinger06}, cold atoms in optical lattices \cite{Jaksch99} 
 and photonic modules \cite{Devitt08}. 
 A quantum gate with an error rate below the threshold required in TEC is 
 within reach of current experimental technology \cite{Benhelm08}. It would be interesting in future work to 
 exploit cluster states of reachable size to implement topologically error-protected quantum algorithms by local measurements.
 
 We acknowledge insightful discussions with M. A. Martin-Delgado, O. G\"uhne. We are grateful to X.-H. Bao for his original idea of the ultra-bright entanglement and to C.-Z. Peng for his idea of reducing high order emission.  We would also like to thank  C. Liu and S. F\"olling for their help in designing the figures.
This work has been supported by the NNSF of China, the CAS, the National Fundamental Research Program 
(under Grant No.  2011CB921300) and NSERC.

\section*{Appendix}

\renewcommand{\thefigure}{S\arabic{figure}}
 \setcounter{figure}{0}
\renewcommand{\theequation}{S.\arabic{equation}}
 \setcounter{equation}{0}
 \renewcommand{\thesection}{S.\Roman{section}}
\setcounter{section}{0}

\section{Topological cluster state quantum computation}

\begin{figure}
\begin{center}
\includegraphics[width=0.95\linewidth]{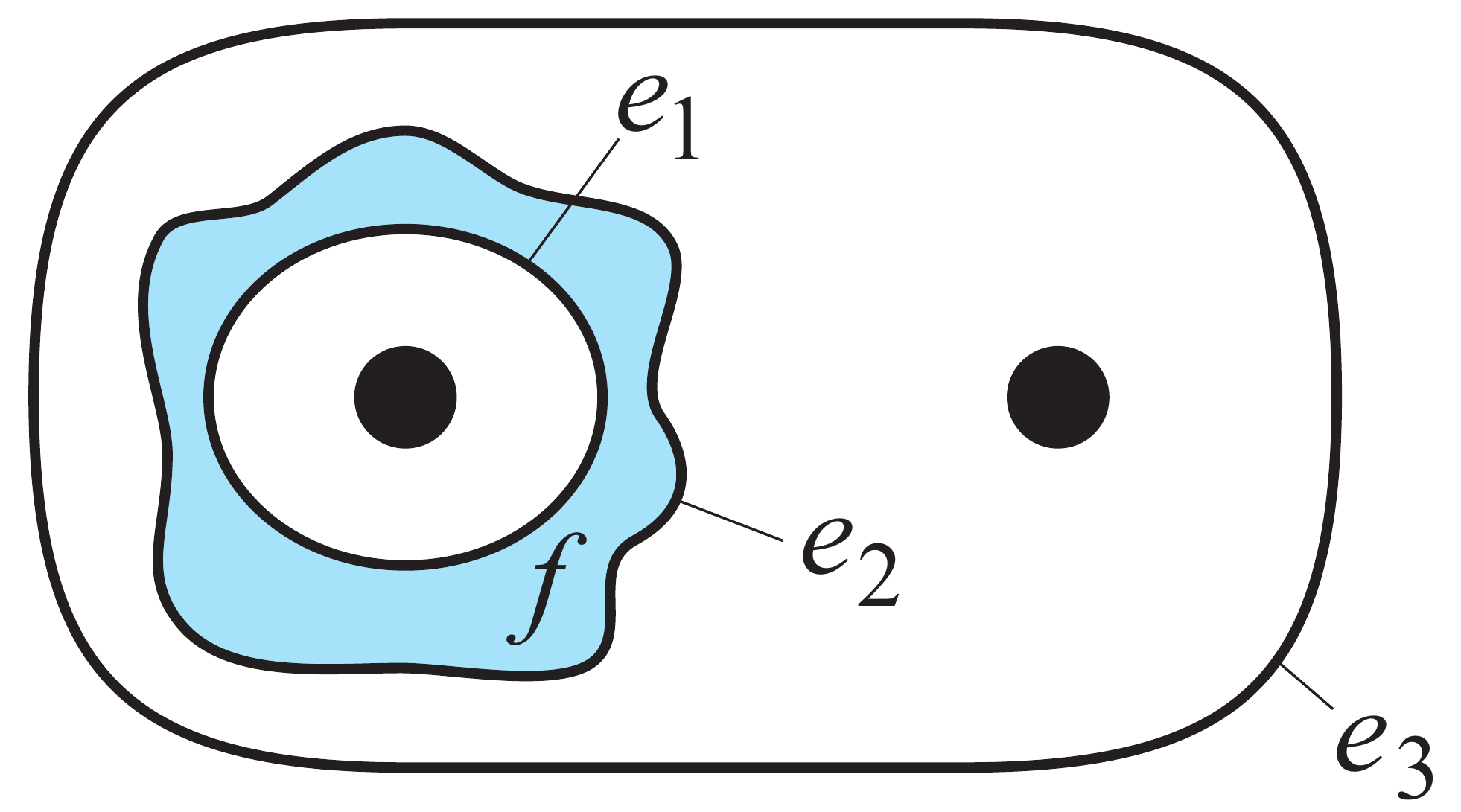}
\caption{\label{Homol}\textbf{ Illustration of homological
equivalence in two dimensions.} Here is a 2D plane with two point
defects ($\bullet$). All three chains, $e_1, e_2$, and $e_3$, are
cycles and thus have no boundary. They are furthermore nontrivial:
each of them is insufficient to be the whole boundary of a surface.
The cycles $e_1$ and $e_2$ are said to be homologically equivalent
because they differ only by the boundary $\partial f$ of the surface
$f$. In contrast, $e_3$ is not equivalent to $e_1$ or $e_2$. In
other words, $e_1$ and $e_2$ can collapse into each other by {\it
smooth} deformation, while $e_3$ cannot be smoothly transformed into
$e_2$ or $e_1$ because of the presence of the right-hand-side
defect. In the topological cluster state quantum computation, errors
are represented by chains. two homologically equivalent error chains
have the same effect on computation.}
\end{center}
\end{figure}

\paragraph{Cluster states and homology.} The topological feature of
error-correction with three-dimensional (3D) cluster states is
homology, which we shall illustrate in 2D for simplicity. Displayed
in Fig.~\ref{Homol} is a 2D plane with two point defects
($\bullet$). The boundary of a surface is defined as the {\it sum}
of all the surrounding chains. For instance, the boundary of the
surface $f$ (shown in blue) is the sum of $e_1$ and $e_2$, denoted
as $\partial f = e_1 + e_2$. Because of the presence of the point
defects, each of the three chains, $e_1$, $e_2$, and $e_3$, is not
sufficient to be the whole boundary of a surface.  Analogously, the
boundary of a chain is defined as the sum of its endpoints. Since
the three chains are cycles, they have no boundary---i.e, $\partial
e_1 = \partial e_2=\partial e_3=0$. The chain $e_2$ can be smoothly
transformed into $e_1$, and vice versa. In other words, $e_1$ and
$e_2$ differ only by the boundary of a surface: $e_2 = e_1 +
\partial f$. We say that $e_1$ and $e_2$ are homologically
equivalent. In contrast, $e_3$ is inequivalent to $e_1$ or $e_2$ due
to the defect on the right-hand side. The homology in higher dimensions is
defined in an analogous way. In 3D, the boundary of a volume is the
sum of all its surrounding surfaces. A closed surface $F$ is said to
have no boundary, i.e., $\partial F =0$. A simple example is the surface
of a sphere. Two surfaces $F$ and $F'$ are homologically equivalent
if they differ only by the boundary of a volume $V$: $F' = F \pm
\partial V$.

In topological cluster state computation, the error correction
scheme only involves local measurements in the $X$ basis, with
outcomes $\lambda =\pm 1$. Computational results are represented by
correlations $R(F)=\prod_{a \in F} \lambda_a$ of these outcomes on a
closed surface $F$---$\partial F=0$. As in any encoding,
error-resilience is brought about by redundancy. A given bit of the
computational result is inferred not only from a particular surface
$F$, but from any one in a huge homology equivalence class. This arises
because two homologically equivalent surfaces $F$ and $F'$ have the
same correlation $R(F)=R(F')$ in the absence of errors \cite{RHG_2}.
As a result, one has $R(\partial V)= 1$ for every volume $V$. An
outcome of -1 then indicates the occurrence of an error in the
volume $V$, and thus $R(\partial V)$ can be used as error syndromes.
We obtain one  bit of such error syndrome per lattice cell; c.f.
Fig.~1a.



The errors have a geometrical interpretation,  too. They correspond 
to 1-chains $e$  \cite{Dennis02}. Again, homology becomes relevant: Two homologically
equivalent error chains $e$ and $e^\prime$ have the same effect on
computation.

In topological error correction with cluster states, the
computational results and the syndromes are contained in
correlations among outcomes of local $X$-measurements. Detecting and
correcting only phase flips of physical qubits is thus sufficient to
correct arbitrary errors. Nevertheless, both bit flip and phase flip
errors are present at the level of logical operations. The qubits in
a 3D cluster state live on the faces and edges of the associated
lattice. Logical phase errors are caused by erroneous measurement of
face qubits, and logical spin flip errors are by erroneous
measurement of edge qubits. For example, the 8-qubit cluster state
$|G_8\rangle$ considered in this experiment has the correlation
$\langle G_8|X_2\otimes X_{2'}|G_8\rangle=1$, in addition to the
four correlations used as error syndromes for face qubits. It can be
derived from a dual complex \cite{RHG_2}, and provides one bit of
(dual) syndrome for the edge qubits of ${\cal{L}}_8$.

\begin{figure}[!t]
\begin{center}
\includegraphics[width=0.95\linewidth]{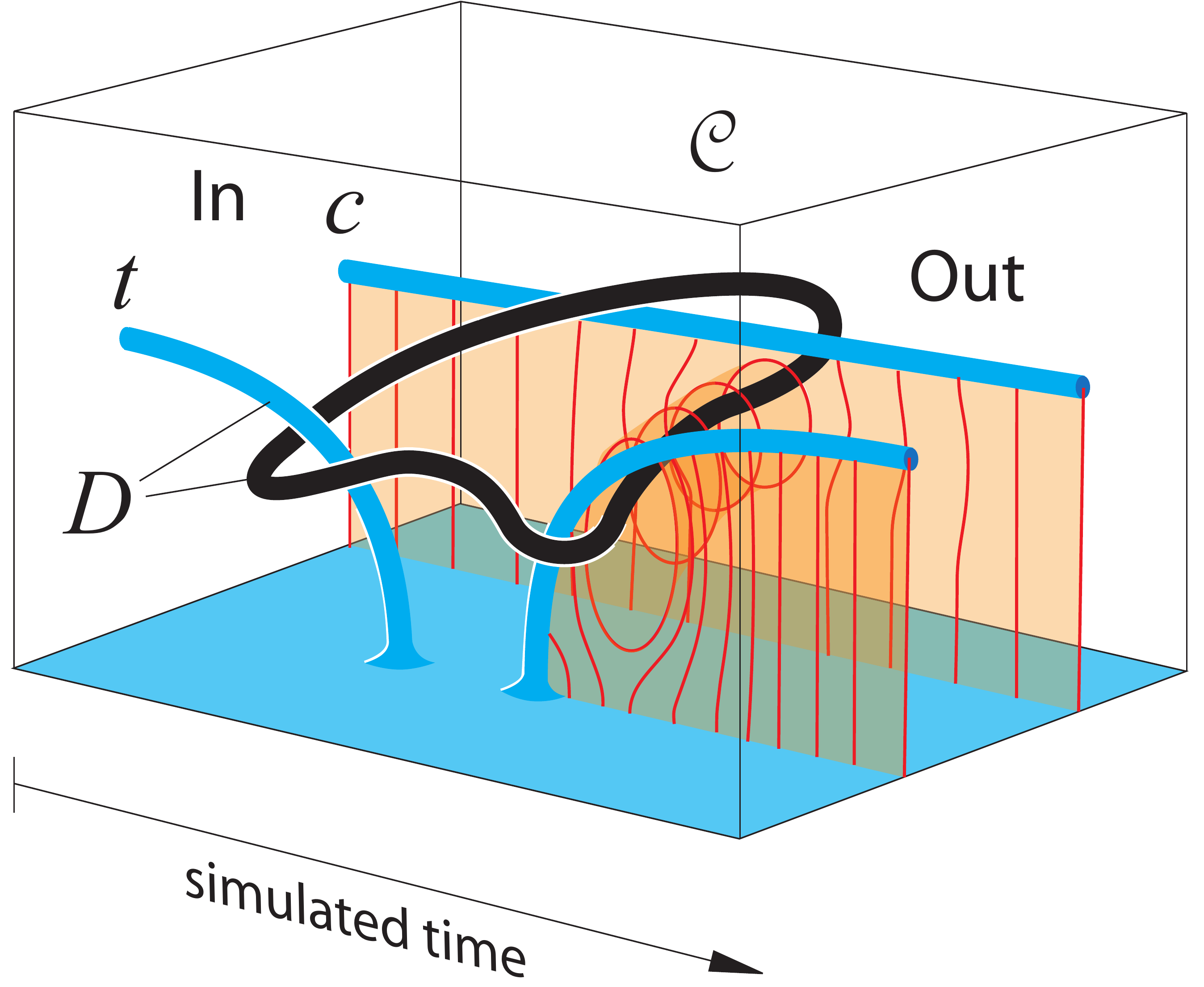}
\caption{\label{Macro}\textbf{Topological gates with 3D cluster
states.} The sub-cluster ${\cal{C}}$ (indicated by the wire frame)
realizes an encoded CNOT gate between control $c$ and target $t$.
The regions $D$ are missing from the cluster. Each line-like such
region supports the world-line of one encoded qubit. One of the four
homologically non-trivial correlation surfaces for the encoded CNOT
is shown in orange.}
\end{center}
\end{figure}

\paragraph{Topologically protected quantum gates.} Topologically protected quantum gates are performed  by
measuring certain regions of qubits in the $Z$ basis, which effectively
removes them. The remaining cluster, whose qubits are to be measured
in the $X$ and $X\pm Y$ basis, thereby attains a non-trivial
topology in which fault-tolerant quantum gates can be encoded.
Fig.~\ref{Macro} shows a macroscopic view of a 3D sub-cluster for
the realization of a topologically protected CNOT gate
\cite{Raus07_2, Raus07d}. Only the topology of the cluster matters,
individual lattice cells are not resolved. The cluster qubits in the
line-like regions $D$ are measured in the $Z$-basis, the remaining
cluster qubits in the $X$-basis.

The fault-tolerance of measurement-based quantum computation with a
3D cluster state can be understood by mapping it to a Kitaev surface
code propagating in time \cite{Raus07_2}. In this picture, a 3D
cluster state consists of many linked toric code surfaces plus extra
qubits for code stabilizer measurement, entangled with these
surfaces. The local measurements in each slice have the effect of
teleporting the encoded state to the subsequent code surface. The
code surfaces can support many encoded qubits because they have
boundary. Encoded gates are implemented by changing the boundary
conditions with time. This process is illustrated in
Fig.~\ref{Macro} for the CNOT gate. Pieces of boundary in the code
surface are created by the intersection of the line-like regions $D$
with surfaces of ``constant time''. The 1-chains displayed in red
represent encoded Pauli operators $\overline{X}$ at a given instant
of simulated time. When propagating forward, an initial operator
$\overline{X}_c$ is converted into $\overline{X}_c \otimes
\overline{X}_t$ as required by conjugation under CNOT.

\paragraph{Further Reading.} For the interested reader we add a few references. The topological
error-correction capability in 3D cluster states is, for the purpose
of establishing long-range entanglement in the presence of noise,
discussed in  \cite{RBH}. How to perform universal fault-tolerant
quantum computation with 3D cluster states is described in
\cite{RHG_2} and in terms of stabilizers in \cite{Fowl08}. In
\cite{Raus07_2}, a mapping from three spatial dimensions to two
spatial dimensions plus time is provided, and the fault-tolerance
threshold is improved to 0.7\%, for both the three and the
two-dimensional version.  The 2D scheme is described solely in terms
of the toric code in \cite{Fowl08b}.

\bigskip

\section{Characterization of the 8-qubit cluster state}

In order to characterize the generated 8-qubit cluster state, we use
entanglement witnesses to verify its genuine multipartite
entanglement \cite{Bou04}. If $\mathcal{W}$ is an observable which
has a positive expectation value on all biseparable states and a
negative expectation value on the generated entangled state, we call
this observable an entanglement witness. With the method introduced
in Ref.~\cite{Otfied07}, the witness is constructed as

\begin{equation}\label{4}
    \mathcal{W}=\frac{1}{2}-\left| \psi  \right\rangle \left\langle
    \psi \right| + \left| \psi'  \right\rangle \left\langle
    \psi' \right|,
\end{equation}
where
\begin{eqnarray}
\left| \psi'  \right\rangle  &=& \frac{1}{2}\left[\left| H\right\rangle^{\otimes 6}
( \left| {VV} \right\rangle-|HH\rangle) \right. \nonumber \\
&& \left. +\left| V\right\rangle^{\otimes 6}
( \left| {HH} \right\rangle+|VV\rangle)\right] \nonumber
\end{eqnarray}
is an orthogonal state of $|\psi\rangle$, that is $\langle \psi|\psi'\rangle=0$.
 
 Then the witness is decomposed into a number of local von Neumann (or projective)
 measurements:
 \begin{widetext}
 \begin{eqnarray}\label{1P}
\mathcal{W}& =& \frac{1}{2} - \left(|\psi\rangle\langle\psi|-|\psi^{\prime}\rangle\langle\psi^{\prime}|\right) \nonumber \\
& =& \frac{1}{2}-\frac{1}{2}\left[(|H\rangle\langle H|^{\otimes 6}-|V\rangle\langle V|^{\otimes 6})_{1-6}\otimes(|H\rangle\langle V|^{\otimes 2}+|V\rangle\langle H|^{\otimes 2})_{78}\right. \nonumber \\
&& \left.+(|H\rangle\langle V|^{\otimes 6}+|V\rangle\langle H|^{\otimes 6})_{1-6}\otimes(|H\rangle\langle H|^{\otimes 2}-|V\rangle\langle V|^{\otimes 2})_{78}\right]\nonumber \\
&=& \frac{1}{2}-\left[\frac{1}{4}\left(|H\rangle\langle H|^{\otimes 6}-|V\rangle\langle V|^{\otimes 6}\right)_{1-6}\otimes\left(\mbox{X}_7\mbox X_8-\mbox Y_7\mbox Y_8\right) +\right. \nonumber \\
&& \left.\frac{1}{12}\left(\sum^{5}_{{k=0}}(-1)^k \mbox M_k^{\otimes 6}\right)_{1-6}\otimes\left(|H\rangle\langle H|^{\otimes 2}-|V\rangle\langle V|^{\otimes 2}\right)_{78}\right],
\end{eqnarray}
\end{widetext}
where $\mbox M_k=\left[\mbox{cos}(\frac{k\pi}{6})\mbox{X}+\mbox{sin}(\frac{k\pi}{6})\mbox{Y}\right]$.
The experimental results are shown in Fig.~4b in the main text, which yields the witness $\langle W\rangle=-0.105\pm0.023$, 
which is negative by 4.5 standard deviations.

\end{document}